\documentclass{ws-procs9x6}
\def\beq{\begin{equation}}
\def\eeq{\end{equation}}
\def\beqn{\begin{eqnarray}}
\def\eeqn{\end{eqnarray}}

\begin{document}

\title{Analysis of SO(2N) Couplings of Spinor and Tensor
Representations in $SU(N)\times U(1)$ Invariant Forms}

\author{ Raza  M.  Syed
}

\address{
Department of Physics, Northeastern University,
Boston, MA 02115-5000, USA}

\maketitle

\abstracts{
A review is given of a recently developed technique for
the analysis of $SO(2N)$ invariant couplings which allows a full
exhibition of the $SU(N)$ invariant content of couplings involving 
the $SO(2N)$ semi-spinors $|\Psi_{\pm}>$ with chiralilty $\pm$  
and tensor representations. We discuss the Basic Theorem used 
in the analysis and then exhibit the technique by illustrative  examples 
for the computation of the trilinear and quartic
couplings for the $SO(10)$ case involving three generations of 16 plets of matter.}

\section{Introduction}
In this paper we give a brief overview of a recently developed technique for
the computation of $SO(2N)$ couplings of spinor and tensor representations in
$SU(N)\times U(1)$ invariant forms. These techniques are then specifically applied
in illustrative examples for the computation of $SO(10)$ invariant couplings when
they are decomposed in $SU(5)\times U(1)$ invariant forms. 
The analysis presented here is of relevance in view of the 
importance of $SO(10)$ as a grand unification group\cite{georgi} of the electroweak 
and the strong interactions. The techniques used are based on the oscillator method\cite{ms,wz,nss}
and developed further in Refs \cite{ns1,ns2,ns3} while 
some related work can be found in Ref.\cite{ab}. 
This paper is thus essentially a brief summary of the works of Refs.\cite{ns1,ns2}.
The outline of the rest of the paper is as follows: In Sec.2 we give a brief discussion
of the $SO(2N)$ algebra, $SO(2N)$ spinor representations for $N$ odd, form of $SO(2N)$ 
invariant couplings, and specialization to the $SO(10)$ case. 
In Sec.3 we give a brief review of the new technique for the evaluation of the 
$SO(2N)$ invariant couplings in terms of $SU(N)\times U(1)$ invariant forms.
In this section we also discuss the Basic Theorem derived in Ref.\cite{ns1}.
In Sec.4 we specialize to the $SO(10)$ case and give illustrative examples of the
computation of cubic couplings in the superpotential and in the Lagrangian.
These involve couplings $16-16-\overline{126}$ in the superpotential, and the couplings
of $16$ plets with $45$ of gauge fields, i.e., the couplings $16-16-45$, in the Lagrangian.
. We also discuss 
a sample of $SO(10)$ invariant quartic couplings. 
\section{Spinor representations of $SO(2N)$; N odd}
 In this section we will discuss
the embedding of $SU(N)$ in $SO(2N)$. Further, we list the general
expressions  for the group invariants formed from different
combinations of $SO(2N)$ spinors. Finally, we specialize the
results to $SO(10)$ grand unification group.
\subsection{${SO(2N)}$ algebra in a  basis}
 The  $2^N$ dimensional spinor, $|{{\Psi}}>$ of $SO(2N)$
 splits into two inequivalent $2^{N-1}$ dimensional semi-spinors,
$|{{\Psi}}_{(\pm)}>$ under the action of the chirality$~$operator:
$|{{\Psi}}_{(\pm)}>=\frac{1}{2}(1 \pm {{\Gamma}}_0)|{{\Psi}}>$
where ${{\Gamma}}_0=i^N
{{\Gamma}}_1{{\Gamma}}_2...{{\Gamma}}_{2N}$ and  further
${{\Gamma}}_0|{{\Psi}}_{(\pm)}>=\pm|{{\Psi}}_{(\pm)}>$. 
Here $\Gamma_{\mu}$ ($\mu=1,2,..,2N$) define a rank $2N$ Clifford algebra 
with $\{\Gamma_{\mu},\Gamma_{\nu}\}=2\delta_{\mu\nu}$, and 
$\Sigma_{\mu\nu}=\frac{1}{2i}[\Gamma_{\mu},\Gamma_{\nu}]$ are the $N(2N-1)$ 
generators of $SO(2N)$. Further, it is convenient to introduce operators 
$b_i$ and $b_i^{\dagger}$ such that $\{b_i,b_j^{\dagger}\}=\delta_{ij}$,
$\{b_i,b_j\}=0=\{b_i^{\dagger},b_j^{\dagger}\}$.
The
semi-spinors, $|{{\Psi}}_{(\pm)}>$ of $SO(2N)$ can be expanded in
terms of reducible antisymmetric $SU(N)$ tensors {${ M},~N$} as follows
\begin{eqnarray}
|{{\Psi}}_{(+)}>= \sum_{p=0,2,..}^{N-1}\frac{1}{p!}{\mathcal
M}^{i_1...i_p}_{(p)}\prod_{q=2,4,..}^p b_{i_q}^{\dagger}|0>,\nonumber\\
|{{\Psi}}_{(-)}>=\sum_{p=1,3,..}^{N}\frac{1}{p!}{\mathcal
N}^{i_1...i_p}_{(p)}\prod_{q=1,3,..}^p b_{i_q}^{\dagger}|0>
\end{eqnarray}
where
the $p$-index tensors can be reduced to $(N-p)$-index tensors as
\begin{eqnarray}
{\mathcal
M}_{(N-p)i_N...i_{p+1}}=\frac{1}{p!}\epsilon_{i_N...i_1}{\mathcal
M}^{i_1...i_p}_{(p)}\nonumber\\
{\mathcal
M}^{i_1...i_p\dagger}_{(p)}={\mathcal M}_{(p)i_p...i_1}^*
\end{eqnarray}

\subsection{$ SO(2N)$ invariant couplings}
Couplings formed from ${\Psi}^{\dagger}$ and ${\Psi}$ are given
 by
 \begin{eqnarray}
 {\mathtt
g}_{ab}<{\Psi}_{(\pm)a}|{\Gamma}_{[\mu_1}.. {\Gamma}_{\mu_p]}
|{\Psi}_{(\mp)b}>{\Phi}_{\mu_1..\mu_p},~p=1,3,..,N
\end{eqnarray}
\begin{eqnarray}
{\mathtt g}_{ab}<{\Psi}_{(\pm)a}|{\Gamma}_{[\mu_1}..
{\Gamma}_{\mu_p]}|{\Psi}_{(\pm)b}>
{\Phi}_{\mu_1..\mu_p},~p=0,2,..,N-1
\end{eqnarray}
where ${\mathtt g}_{ab}$ is a
coupling constant where $a$ and $b$ are family indices.
${\Phi}_{\mu_1\mu_2...\mu_p}$ is a real antisymmetric tensor of
$SO(2N)$ with dimensionality $2N \choose p$. For $p=N$
\begin{eqnarray}
{\Phi}_{\mu_1..\mu_N}
 ={\overline{{\Delta}}}_{\mu_1..\mu_N}+
{\Delta}_{\mu_1..\mu_N},\nonumber\\
 {{\overline
{\Delta}_{\mu_1..\mu_N}}\choose {\Delta}_{\mu_1..\mu_N}}=
\pm\frac{i}{N!}{\mathbf{\epsilon}}_{{\mu_{1}..\mu_{N}\nu_{1}..\nu_{N}}}{{\overline
{\Delta}_{\nu_1..\nu_N}}\choose {\Delta}_{\nu_1..\nu_N}}. 
\end{eqnarray}
Both
${\overline{{\Delta}}}$ and ${\Delta}$ have dimensionality
$\frac{1}{2}{2N \choose N}$. The symmetry factor in the exchange
of identical ${ \Psi}_{(.)}$ is ${\mathtt g}_{ab}=(-1)^{\frac
{1}{2}{p(p-1)}}{\mathtt g}_{ba}$. Couplings formed from
${\Psi}^{\bf{T}}$ and ${\Psi}$ are given by
\begin{eqnarray}
{\mathtt f}_{ab}<{\Psi}_{(\pm)a}^*| B{\Gamma}_{[\mu_1}..
{\Gamma}_{\mu_p]} {|{\Psi}_{(\pm)b>}
|{\Psi}_{(\mp)b}>}{\Phi}_{\mu_1..\mu_p},~p=1,3,..,N
\end{eqnarray}
\begin{eqnarray}
{\mathtt f}_{ab}<{\Psi}_{(\pm)a}^*|B{\Gamma}_{[\mu_1}..
{\Gamma}_{\mu_p]}|{\Psi}_{(\mp)b}>{\Phi}_{\mu_1..\mu_p},~p=0,2,..,N-1
\end{eqnarray}
where
$B=\prod_{\mu =1}^N{\Gamma}_{2\mu -1}$ is the $SO(2N)$ charge
conjugation operator and satisfies the relation
${\acute{\Sigma}}_{\mu\nu}^{\bf T}{\acute B}=-
{\acute{\Sigma}}_{\mu\nu}{\acute B}$ where $~\acute{  }~$
indicates a $2^N\times 2^N$ matrix representation. The symmetry
factor ${\mathtt f}_{ab}=(-1)^{\frac{1}{2}(N-p)(N-p-1)}{\mathtt
f}_{ba}$.
\subsection{
Specialization to $SO(10)$ gauge group } 
We consider now the special case of $SO(10)$ where 
${\Psi}_{(+)}\sim 16
$,${\Psi}_{(-)}\sim\overline {16}$ under $SO(10)\supset
SU(5)\otimes U(1)$. Here 
$16\supset [1]\oplus[\overline 5]\oplus [10]$,
$\overline {16}\supset [1]\oplus [5]\oplus[\overline{10}]$ and 
$16\otimes 16=10_s\oplus 120_{as}\oplus 126_s$ while
 $16\otimes\overline {16}=1\oplus 45\oplus 210$.
In terms of their oscillator modes
\begin{eqnarray}
 |{\Psi}_{(+)a}>=|0>{\mathcal M}_a+\frac{1}{2}b_i^{\dagger}b_j^{\dagger}|0>{\mathcal M}_a^{ij}
+\frac{1}{24}\epsilon^{ijklm}b_j^{\dagger}
b_k^{\dagger}b_l^{\dagger}b_m^{\dagger}|0>{\mathcal
M}_{ai}\nonumber\\
 |{\Psi}_{(-)b}>=
\frac{1}{12}\epsilon^{ijklm}b_k^{\dagger}b_l^{\dagger}b_m^{\dagger}|0>{\mathcal
N}_{bij}+b_1^{\dagger}b_2^{\dagger}
b_3^{\dagger}b_4^{\dagger}b_5^{\dagger}|0>{\mathcal
N}_b+b_i^{\dagger}|0>{\mathcal N}_b^i
\end{eqnarray}
where
\begin{eqnarray}
 {\mathcal
M}_a=\nu_{La}^c,~{\mathcal
M}_{a\alpha}=D_{La\alpha}^c,~{\mathcal
M}_a^{\alpha\beta}=\epsilon^{\alpha\beta\gamma}U_{La\gamma}^c,~
{\mathcal M}_{a4}=E_{La}^{^{-}},\nonumber\\
{\mathcal M}_a^{4\alpha}=U_{La\alpha},~{\mathcal
M}_{a5}=\nu_{La},~ {\mathcal M}_a^{45}=E_{La}^{^{+}},~{\mathcal
M}_a^{5\alpha}=D_{La\alpha}
\end{eqnarray}
and where $\alpha,\beta,\gamma$ are the color indices. Cubic couplings 
in the superpotential involving ${\Psi}_{(\pm)}$ are
\begin{eqnarray}
{\mathsf W}^{(1)}={
f}^{(1)}_{ab}<\widehat{{\Psi}}_{(\pm)a}^*|B|\widehat{{\Psi}}_{(\mp)b}>{\Phi}\nonumber\\
{\mathsf W}^{(45)}=\frac{1}{2!}{
f}^{(45)}_{ab}<\widehat{{\Psi}}_{(\pm)a}^*|B
{\bf{\Sigma}}_{\mu\nu}|\widehat{{\Psi}}_{(\mp)b}>{\Phi}_{\mu\nu}\nonumber\\
 {\mathsf W}^{(210)}=\frac{1}{4!} {f}^{(210)}_{ab}<\widehat{{\Psi}}_{(\pm)a}^*|B
{\Gamma}_{[\mu}{\Gamma}_{\nu}{\Gamma}_{\rho} {\Gamma}_{\lambda]}
|\widehat{{\Psi}}_{(\mp)b}>{\Phi}_{\mu\nu\rho\lambda}\nonumber\\
{\mathsf W}^{(10)}={
f}^{(10)}_{ab}<\widehat{{\Psi}}_{(\pm)a}^*|B{\Gamma}_{\mu}|\widehat{{\Psi}}_{(\pm)b}>{\Phi}_{\mu}\nonumber\\
{\mathsf W}^{(120)}=\frac{1}{3!}{
f}^{(120)}_{ab}<\widehat{{\Psi}}_{(\pm)a}^*|B
{\Gamma}_{[\mu}{\Gamma}_{\nu}
{\Gamma}_{\lambda]}|\widehat{{\Psi}}_{(\pm)b}>{\Phi}_{\mu\nu\lambda}\nonumber\\
 {\mathsf W}^{(126,\overline{126})}=\frac{1}{5!}{f}^{(126,\overline{126})}_{ab}<\widehat{{\Psi}}_{(\pm)a}^*|B
{\Gamma}_{[\mu}{\Gamma}_{\nu}{\Gamma}_{\rho}{\Gamma}_{\sigma}{\Gamma}_{\lambda]}|\widehat{{\Psi}}_{(\pm)b}>{\overline{{\bf{\Delta}}}_{\mu\nu\rho\sigma\lambda}\choose{\bf{\Delta}}_{\mu\nu\rho\sigma\lambda}}
\end{eqnarray}
The semi-spinor $\Psi_{(\pm)}$ with  $~\widehat{  }~$ stands for a 
chiral superfield and $SO(10)$ charge conjugation operator is $B=
-i\prod_{k=1}^5 (b_k-b_k^{\dagger})$.
 Couplings in the Lagrangian have the form
\begin{eqnarray}
{\mathsf L}^{(1)}={
g}^{(1)}_{ab}<{{\Psi}}_{(\pm)a}|\gamma^0\gamma^A|{{\Psi}}_{(\pm)b}>{\Phi}_{A}\nonumber\\
{\mathsf L}^{(45)}=\frac{1}{2!}{
g}^{(45)}_{ab}<{{\Psi}}_{(\pm)a}|\gamma^0\gamma^A
{\bf{\Sigma}}_{\mu\nu}|{{\Psi}}_{(\pm)b}>{\Phi}_{A\mu\nu}\nonumber\\
{\mathsf L}^{(210)}=\frac{1}{4!} {
g}^{(210)}_{ab}<{{\Psi}}_{(\pm)a}|\gamma^0\gamma^A
{\Gamma}_{[\mu}{\Gamma}_{\nu}{\Gamma}_{\rho} {\Gamma}_{\lambda]}
|{{\Psi}}_{(\pm)b}>{\Phi}_{A\mu\nu\rho\lambda}
\end{eqnarray}
where $A$ stands for the Lorentz index.

\section{Technique for Evaluation of $SO(2N)$ invariants }
Here we review the recently developed technique\cite{ns1,ns2}
 for the analysis of $SO(2N)$ invariant
couplings which allows a full exhibition of the $SU(N)$ invariant
 content of the spinor and tensor
representations. The technique utilizes a basis consisting of a
specific set of reducible $SU(N)$ tensors in terms of which the
SO(2N) invariant couplings have a simple expansion.

\subsection{Specific set of $ SU(N)$ reducible tensors}
We begin with the observation that the natural basis for the
expansion of the $SO(2N)$ vertex is in terms of a specific set of
$SU(N)$ reducible tensors, ${\Phi}_{c_k}$ and ${\Phi}_{\overline
c_k}$ which we define as
$A^k\equiv{\Phi}_{c_k}\equiv{\Phi}_{2k}+i{\Phi}_{2k-1},~
A_k\equiv{\Phi}_{\overline c_k}\equiv{\Phi}_{2k}-i{\Phi}_{2k-1}$.
This can be extended immediately
 to define the quantity $\Phi_{c_ic_j\bar c_k..}$
with an arbitrary number of unbarred and barred indices where each
 $c$ index can be expanded out so that
$A^iA^jA_k...={\Phi}_{c_ic_j\overline
c_k...}={\Phi}_{2ic_j\overline c_k...}+i{\Phi}_{2i-1c_j\overline
 c_k...}~$etc..
 Thus, for example, the quantity  $\Phi_{c_ic_j\overline c_k...c_N}$ is a sum of
 $2^N$ terms gotten by expanding all the c indices.
$\Phi_{c_ic_j\overline c_k...c_n}$ is completely anti-symmetric in
the interchange of its c indices whether unbarred or barred:
 ${\Phi}_{c_i\overline c_jc_k...\overline c_n}=-{\Phi}_{c_k\overline c_jc_i...\overline c_n}$.
Further,
 $ {\Phi}^*_{c_i\overline c_jc_k...\overline c_n}={\Phi}_{\overline c_ic_j\overline
  c_k...c_n}$ etc.. We now make the observation\cite{ns1} that the object
$\Phi_{c_ic_j\overline c_k...c_n}$ transforms like a reducible
representation of SU(N). Thus if we are able to compute the SO(2N)
invariant couplings
 in terms of these reducible tensors of SU(N) then
there remains only the further step of decomposing the reducible
 tensors into their irreducible parts.

\subsection{Basic Theorem to evaluate an $ SO(2N)$ vertex}
A result essential to our analysis is the Basic Theorem\cite{ns1} which states that an
 $SO(2N)$ vertex ${\Gamma}_{\mu}{\Gamma}_{\nu}{\Gamma}_{\lambda}...{\Gamma}_{\sigma}{\Phi}_{\mu\nu\lambda
 ...{\bf{\sigma}}}$ can be expanded in the following fashion
\begin{eqnarray}
{\Gamma}_{\mu}{\Gamma}_{\nu}{\Gamma}_{\lambda}...{\Gamma}_{\sigma}{\Phi}_{\mu\nu\lambda
...{{\sigma}}}=
b_i^{\dagger}b_j^{\dagger}b_k^{\dagger}...b_n^{\dagger}{\Phi}_{c_ic_jc_k...c_n}~~~~~~~~~\nonumber\\
+(b_i b_j^{\dagger}b_k^{\dagger}...b_n^{\dagger}{\Phi}_{\overline
c_ic_jc_k...c_n}+ \textnormal{perms})
 +(b_i
b_jb_k^{\dagger}...b_n^{\dagger}{\Phi}_{\overline c_i\overline
c_jc_{k}...c_n}+~\textnormal{perms})+ ...~~~\nonumber\\
+(b_ib_jb_k...b_{n-1}b_n^{\dagger}{\Phi}_{\overline c_i\overline
c_j \overline c_k...\overline c_{n-1}c_n}+~\textnormal{perms})~~~~~~~~\nonumber\\
+b_ib_jb_k...b_n{\Phi}_{\overline c_i\overline c_j\overline
c_k...\overline c_n}~~~~~~~~
\end{eqnarray}
The result of Eq.(12) is very useful in the computation of $SO(10)$ invariant couplings.

\section{Cubic  Couplings of ${\bf SO(10)}$}
In this section we give illustrative examples  of some
$SO(10)$ trilinear couplings in their $SU(5)$ decomposed form.
 These illustrative examples consist of $16-16-45$ couplings in the Lagrangian and
 the $16-16-\overline{126}$ coupling in the superpotential.

\subsection{$16\otimes 16\otimes 45$ coupling in the Lagrangian}
The interaction Lagrangian of the $45$ of gauge fields with the 16-plet of $SO(10)$ spinor
$|\Psi_{(+)}>$ is given by 
\begin{eqnarray}
{\mathsf
L}^{(45)}=\frac{1}{i}\frac{1}{2!}g^{(45)}_{ab}<\Psi_{(+)a}|\gamma^0\gamma^A
\Sigma_{\mu\nu}|\Psi_{(+)b}>\Phi_{A\mu\nu}.
\end{eqnarray}
Expansion of the vertex gives
 \begin{eqnarray}
\Sigma_{\mu\nu}\Phi_{\mu\nu}=\frac{1}{i}(b_ib_j \Phi_{\overline
c_i\overline c_j}+b_i^{\dagger}b_j^{\dagger}
\Phi_{c_ic_j}+2b_i^{\dagger}b_j\Phi_{c_i\overline c_j}-
\Phi_{c_n\overline c_n}).
\end{eqnarray}
The $45$ of $SO(10)$ decomposes under
$SU(5)$ as
 $45 \supset 1({\mathsf g})\oplus 10({\mathsf g}^{ij})\oplus{\overline{10}}({\mathsf
g}_{ij})\oplus {\overline {24}}({\mathsf g}^i_{j})$ where
\begin{eqnarray}
\Phi_{c_n\overline c_n}={\mathsf g},~~\Phi_{c_i\overline
c_j}={\mathsf g}_{j}^i+ \frac{1}{5}\delta_j^i{\mathsf
g}\nonumber\\
\Phi_{c_ic_j}={\mathsf g}^{ij},~~\Phi_{\overline c_i\overline
c_j}={\mathsf g}_{ij}.
\end{eqnarray}
The normalized $SU(5)$ gauge fields are
\begin{eqnarray}
{\mathsf g}_A=2\sqrt 5 {\mathsf G}_A,~~{\mathsf g}_{Aij}=\sqrt 2
{\mathsf G} _{Aij}\nonumber\\
 {\mathsf g}_A^{ij}=\sqrt 2 {\mathsf
G}_A^{ij},~~{\mathsf g}_{Aj}^i=\sqrt{2} {\mathsf G}_{Aj}^i
\end{eqnarray}
In terms of the redefined fields, the kinetic energy of the 45-plet
takes the form
\begin{eqnarray}
-\frac{1}{4}{ F}_{\mu\nu}^ {AB}{ F}_{AB\mu\nu}=-\frac{1}{2}{~
G}_{AB}{G}^{AB\dagger} -\frac{1}{2!} \frac{1}{2}{G}^{ABij}{
G}_{AB}^{ij\dagger} -\frac{1}{4}{ G}_j^{ABi}{ G}_{ABi}^{j}
\end{eqnarray}
where ${ F}_{\mu\nu}^{AB}$ is the 45 of SO(10) field strength
tensor
\begin{eqnarray}
 {\mathsf
L}^{(45)}=g^{(45)}_{ab}[\sqrt 5(-\frac{3}{5}\overline {\mathcal
M}_a^i\gamma^A{\mathcal M}_{bi}+\frac{1}{10}\overline {\mathcal
M}_{aij} \gamma^A{\mathcal M}_b^{ij}+ \overline{\mathcal
M}_a\gamma^A{\mathcal M}_b){\mathsf G}_{A}\nonumber\\
 +{\frac{1}
{\sqrt 2}}(\overline {\mathcal M}_a \gamma^{A} {\mathcal M}_b^{lm}
+{\frac{1}{2}}\epsilon^{ijklm}\overline {\mathcal
M}_{aij}\gamma^A{\mathcal M}_{bk}) {\mathsf G}_{Alm}\nonumber\\
-\frac{1}{\sqrt 2}(\overline {\mathcal M}_{alm}\gamma^A{\mathcal
M}_b+\frac{1}{2}\epsilon_{ijklm} \overline {\mathcal
M}_a^i\gamma^A{\mathcal M}_b^{jk}){\mathsf
G}_A^{lm}\nonumber\\
+\sqrt{2}(\overline {\mathcal M}_{aik}\gamma^A{\mathcal
M}_b^{kj}+\overline {\mathcal M}_a^j\gamma^A{\mathcal
M}_{bi}){\mathsf G}_{Aj}^i]
\end{eqnarray}
 The barred matter fields are defined
so that $~\overline {\mathcal M}_{ij}={\mathcal
M}_{ij}^{\dagger}\gamma ^0$.
\subsection{${16}\otimes {16} \otimes$ ${\overline {126}}$ coupling in the
superpotential}
The 
${\overline {126}}$ of $SO(10)$ decomposes under
$SU(5)$ as ${\overline {126}}\supset 1({\mathsf H})\oplus
5({\mathsf H}^{i})\oplus \overline{10}({\mathsf H}_{ij})\oplus
15({\mathsf H}_{(S)}^{ij})\oplus \overline{45}({\mathsf
H}_{ij}^k)\oplus 50({\mathsf H}^{ij}_{kl})$. 
Utilizing the Basic Theorem the
result for the Yukawa coupling involving ${\overline {126}}$ of
Higgs is as follows
\begin{eqnarray}
{\mathsf
W}^{(\overline{126})}=\frac{1}{5!}{f}^{(\overline{126})}_{{a}{b}}<\widehat{\Psi}^{*}_{(+){a}}|B\Gamma_{[\mu}\Gamma_{\nu}\Gamma_{\rho}
\Gamma_{\lambda}\Gamma_{\sigma]}
|\widehat{\Psi}_{(+){b}}>\overline{\Delta}_{\mu\nu\rho\lambda\sigma}\nonumber\\
=i\sqrt{\frac{2}{15}}{f}^{(\overline{126})(+)}_{ab}[
-\sqrt{2}\widehat{\mathcal M}_{a}^{\bf T}\widehat{\mathcal
M}_{b}{\mathsf H}+\widehat{\mathcal M}_{a}^{\bf
T}\widehat{\mathcal M}_{b}^{ij}{\mathsf H}_{ij}\nonumber\\
-\widehat{\mathcal M}_{ai}^{\bf T}\widehat{\mathcal
M}_{bj}{\mathsf H}_{(S)}^{ij}+\widehat{\mathcal M}_{a}^{ij\bf
T}\widehat{\mathcal M}_{bk}{\mathsf H}_{ij}^{k}\nonumber\\
-\frac{1}{12\sqrt{2}}\epsilon_{ijklm}\widehat{\mathcal
M}_{a}^{ij\bf
T}\widehat{\mathcal M}_{b}^{rs}{\mathsf H}^{klm}_{rs}\nonumber\\
-\sqrt{3}\left(\widehat{\mathcal M}_{a}^{\bf T}\widehat{\mathcal
M}_{bm} +\frac{1}{24}\epsilon_{ijklm}\widehat{\mathcal
M}_{a}^{ij\bf T}\widehat{\mathcal M}_{b}^{kl}\right){\mathsf H}^m]
\end{eqnarray}
 where
$f^{({\overline {126}})(+)}_{ab}=\frac{1}{2}(f^{({\overline
{126}})}_{ab}+f^{({\overline {126}})}_{ba})$.

\section{Quartic Couplings of $SO(10)$}
The technique discussed in Secs.3-4 can be used to compute the quartic couplings.
For illustrative purposes we consider the simplest example with the 
 superpotenial,
\begin{eqnarray}
 {\mathsf
W}=\frac{1}{2}\Phi_{U}{\bf M}^{(1)}_{{U}{U}'}\Phi_{{
U}'}
+f^{(1)}_{ab}<\Psi^{*}_{(-)a}|B |\Psi_{(+)b}>l_{ U}^{(1)}\Phi_{ U} +..
\end{eqnarray}
where the indices ${U},~{ U}'$  run over several Higgs
representations of the same kind. ${\bf M}^{(1)}$
represents the mass matrices and $f^{(1)}$
 are constants. We now eliminate $\Phi_{ U}$  using the F-flatness
condition: $\frac{\partial {\mathsf W}}{\partial \Phi_{U}}=0$.
This leads to\cite{ns2}
\begin{eqnarray}
 {\mathsf
W}^{(16\times {\overline {16}})_{1}(16\times \overline {16})_{1}}
= 2\lambda_{ab,cd}^{^{(1)}}
<\widehat{\Psi}^*_{(-)a}|B|\widehat{\Psi}_{(+)b}><\widehat{\Psi}^*_{(-)c}|B|
\widehat{\Psi}_{(+)d}>\nonumber\\
=\frac{1}{2}\lambda_{ab,cd}^{^{(1)}} [-\widehat{\mathcal
N}_{aij}^{\bf{T}}\widehat {\mathcal M}^{ij}_{b}\widehat{\mathcal
N}_{ckl}^{\bf{T}}\widehat {\mathcal M}^{kl}_{d}+4\widehat{\mathcal
N}^{i\bf{T}}_{a}\widehat {\mathcal M}_{bi}\widehat{\mathcal
N}_{cjk}^{\bf{T}}\widehat {\mathcal M}^{jk}_{d}\nonumber\\
-4\widehat{\mathcal N}^{i\bf{T}}_{a}\widehat {\mathcal
M}_{bi}\widehat{\mathcal N}^{j\bf{T}}_{c} \widehat {\mathcal
M}_{dj} +4\widehat{\mathcal N}^{\bf{T}}_{a}\widehat {\mathcal
M}_{b}\widehat{\mathcal N}^{\bf{T}}_{cij}\widehat {\mathcal
M}^{ij}_{d} \nonumber\\
-8\widehat{\mathcal N}^{\bf{T}}_{a}\widehat
{\mathcal M}_{b}\widehat{\mathcal N}^{i\bf{T}}_{c}\widehat
{\mathcal M}_{di} -4\widehat{\mathcal N}_{a}^{\bf{T}}\widehat
{\mathcal M}_{b}\widehat{\mathcal N}_{c}^{\bf{T}}\widehat
{\mathcal M}_{d}]
\end{eqnarray}
where
\begin{eqnarray}
 \lambda_{ab,cd}^{^{(1)}}=
f_{ab}^{(1)}f_{cd}^{(1)}l_{ U}^{(1)}\left[\widetilde{{\bf
M}}^{(1)}\left\{{\bf M}^{(1)}\widetilde{{\bf M}}^{(1)}
-\bf{1}\right\}
\right]_{{U}{U}'}l_{ U'}^{(1)}\nonumber\\
\widetilde{{\bf M}}^{(.)}=\left[{\bf M}^{(.)}+\left({\bf
M}^{(.)}\right)^{\bf {T}}\right]^{-1}
\end{eqnarray}
Similarly a complete determination of $(16\times {\overline {16}})_{45} 
(16\times {\overline {16}})_{45}$ and of 
$(16\times {\overline {16}})_{210} 
(16\times {\overline {16}})_{210}$ can be given\cite{ns2}.

\section{Conclusion}
In this paper we have given a brief overview of the $SO(2N)$ ($N$ odd) invariant
couplings. In Sec.2 we gave a brief summary of some of the salient features of 
$SO(2N)$ algebra in terms of oscillator modes. We exhibited the form of $SO(2N)$ 
invariant cubic couplings and then specialized these results to the $SO(10)$ case.
In Sec.3 we introduced a basis involving reducible $SU(N)$ tensors in terms of which
$SO(2N)$ invariant couplings have a simple expansion. This result is codified in 
the so called Basic Theorem which is stated at the end of Sec.3. In Sec.4 we used the
Basic Theorem to decompose $SO(10)$ invariant cubic couplings in terms of $SU(5)\times U(1)$ 
invariant forms. An application of the Basic Theorem was given through two illustrative
examples involving the $16$ plet spinor representations and the tensor representations,
45 and $\overline{126}$. The analysis presented here should be of interest to model 
builders using $SO(2N)$ (N odd) type gauge groups.

\section{Acknowledgments}
This paper is dedicated to Prof. Pran Nath on the occasion of his
sixty fifth birthday and for proceedings of NathFest. 
This research was supported in part by NSF grant
PHY-0139967.


\begin{thebibliography}{99}



\bibitem{georgi}
H. Georgi, in Particles and Fields (edited by C.E. Carlson), A.I.P.,
1975; H. Fritzch and P. Minkowski, Ann. Phys. {\bf 93}(1975)193.

\bibitem{ms}
R.N. Mohapatra and B. Sakita, Phys. Rev.{\bf D21}(1980)1062.

\bibitem{wz}
 F. Wilczek and A. Zee, Phys. Rev. {\bf D25}(1982)553.

 \bibitem{nss}
   S. Nandi, A.
Stern, E.C.G. Sudarshan, Phys. Rev. {\bf D26}(1982)1653

\bibitem{ab}
For alternative approaches see, 
X.~G.~He and S.~Meljanac,
Phys.\ Rev.\ D {\bf 41} (1990) 1620.
;
G.~W.~Anderson and T.~Blazek,
J.\ Math.\ Phys.\  {\bf 41}, 4808 (2000)
[arXiv:hep-ph/9912365].
:
C.~S.~Aulakh and A.~Girdhar,
arXiv:hep-ph/0204097.


\bibitem{ns1}
P.~Nath and R.~M.~Syed,
Phys.\ Lett.\ B {\bf 506}, 68 (2001)


\bibitem{ns2}
P.~Nath and R.~M.~Syed,
Nucl.\ Phys.\ B {\bf 618}, 138 (2001)


\bibitem{ns3}
P.~Nath and R.~M.~Syed,
Nucl.\ Phys.\ B {\bf 676}, 64 (2004)


\end{thebibliography}
\end{document}